\documentclass[twoside]{article}
\usepackage{my_qic}
\usepackage{graphicx}

\newcommand{\<}{\langle}
\renewcommand{\>}{\rangle}

\newcommand{\be}{\begin{equation}}
\newcommand{\ee}{\end{equation}}

\begin{document}

\runninghead{Secure assisted quantum computation}
            {Andrew M.\ Childs}

\normalsize\textlineskip
\thispagestyle{empty}
\setcounter{page}{1}

\vspace*{0.88truein}

\alphfootnote

\fpage{1}

\centerline{\bf
SECURE ASSISTED QUANTUM COMPUTATION}
\vspace*{0.37truein}
\centerline{\footnotesize
ANDREW M.\ CHILDS}
\vspace*{0.015truein}
\centerline{\footnotesize\it Center for Theoretical Physics}
\baselineskip=10pt
\centerline{\footnotesize\it Massachusetts Institute of Technology}
\centerline{\footnotesize\it Cambridge, Massachusetts 02139, USA}
\vspace*{0.225truein}

\abstracts{
Suppose Alice wants to perform some computation that could be done
quickly on a quantum computer, but she cannot do universal quantum
computation.  Bob can do universal quantum computation and claims he
is willing to help, but Alice wants to be sure that Bob cannot learn
her input, the result of her calculation, or perhaps even the function
she is trying to compute.  We describe a simple, efficient protocol by
which Bob can help Alice perform the computation, but there is no way
for him to learn anything about it.  We also discuss techniques for
Alice to detect whether Bob is honestly helping her or if he is
introducing errors.
}{}{}

\vspace*{1pt}\textlineskip

\section{Introduction}
\label{sec:intro}

The idea of processing information stored in quantum states has
spawned numerous cryptographic applications.  A few examples include
quantum key distribution \cite{BB84}, which allows remote parties to
securely establish a shared list of random numbers; a fast quantum
algorithm for factoring \cite{Sho94}, which can be used to break
certain classical cryptosystems; quantum secret sharing \cite{CGL99},
by which a secret quantum state can be divided among several parties;
quantum data hiding \cite{DLT02,EW02,DHT02}, which offers an
information theoretically secure way of sharing a classical secret;
quantum digital signatures \cite{GC01}, which can be used to
authenticate documents; and secure quantum channels
\cite{BBCJPW93,AMTW00,BR00,Leu01}, which allow secure transmission of
quantum states.

But there are also a number of negative results about the possible
cryptographic applications of quantum information, such as the
impossibility of an unconditionally secure quantum protocol for bit
commitment \cite{LC97,May97}.  A related result is the impossibility
of ``secure two-party computation,'' in which two parties collaborate
to compute a function without revealing their inputs \cite{Lo97}
(although in general, secure {\em multi}-party quantum computation is
possible \cite{Cha99,Smi01}).  However, this does not rule out all
forms of collaborative computation by two parties: for example, what
if one of the parties wishes to assist the other, with no possibility
of learning the input or output of the computation?  We will show that
this kind of two-party computation {\em can} be done securely.

More precisely, the problem we will consider is the following: imagine
that Alice would like to perform a quantum computation in secret, but
although she can do some basic quantum gates, she does not have a
full-fledged quantum computer.  Bob, who runs a company that sells
time on its quantum supercomputer, would like to supply Alice with the
resources she needs to perform her computation.  But Alice does not
trust Bob; she needs to be absolutely sure that he cannot learn
anything about her computation, i.e., that the state Bob sees at any
time is the same, independent of her actual quantum state.  Can they
carry out an unconditionally secure protocol by which Bob can assist
Alice?  In this paper, we describe protocols that answer this question
in the affirmative.

To fully specify our question, we must decide exactly what resources
are allowed.  Three kinds of resources must be specified: operations
available to Alice, operations available to Bob, and ways in which
they can communicate.  We will always allow Bob to do universal
quantum computation and make arbitrary quantum measurements, and we
will allow bidirectional quantum communication.  There are many
possible restrictions on Alice's resources that might be of interest,
but we will choose the most restrictive set under which Bob can help
her do universal quantum computation.  We will allow Alice to store
quantum states and to route her qubits (i.e., perform the {\sc swap}
gate), but we will suppose that the only nontrivial gates she can
perform are the {\em Pauli gates}, 
\be
  X := \left[\matrix{0 & 1 \cr 1 &  0}\right]\,, \quad
  Z := \left[\matrix{1 & 0 \cr 0 & -1}\right]
\ee
(with which she can perform their product $XZ$, etc.).  This gate set
fails to be universal in two important ways.  First, Alice cannot
perform any interactions between qubits.  Second, the single-qubit
gates she can perform are restricted to the discrete set
$\{I,X,Z,XZ\}$ (up to an overall phase\footnote{For simplicity, we
identify operators that differ by an overall phase, since such phases
are irrelevant in all the situations we consider.  Equivalently, we
could multiply all operators by a phase so that they have unit
determinant.}~), which forms a group (the {\em Pauli group}) under
multiplication.  In addition to the restrictions on her gates, we will
suppose that Alice can only prepare the $|0\>$ state, and she cannot
perform measurements.  Alice also must be able to generate random
classical bits (say, by flipping unbiased coins) and to perform Pauli
gates conditioned on the values of these classical bits.

Note that a related question has been considered in the classical
setting.  As we argue in Section~\ref{sec:classical}, there is no
reasonable restriction on Alice's gate set under which she can carry
out a classical protocol analogous to our quantum protocols.  However,
one can instead assume Alice has the ability to perform
polynomial-time computation and ask whether she can securely receieve
help from an arbitrarily powerful Bob.  For certain particular
problems this is the case \cite{Fei86,FF93}, whereas for other
problems (in particular, NP-hard ones), it is not \cite{AFK89,FF93}.
In contrast, our protocols make no computational assumptions, and will
apply to {\em any} quantum circuit known to Alice, not just the
computation of a particular function.

All of the protocols we present are applications of a quantum version
of the Vernam cipher \cite{Ver26}, also known as private key
encryption or the one-time pad.  The classical Vernam cipher works as
follows: suppose Alice wishes to securely send a bit $b$ to Charlie.
She and Charlie share another bit $k$, the {\em key}, that is randomly
chosen to be either $0$ or $1$, each with probability $1 \over 2$.
Alice computes the message bit $m=b \oplus k$, where $\oplus$ denotes
addition modulo two, and sends it to Charlie.  Since he knows $k$, he
can compute $b = m \oplus k$.  However, an eavesdropper (whom we shall
call Eve) who does not know $k$ can learn nothing about $b$ since $b$
and $m$ have zero mutual information.  To send multiple bits, Alice
and Charlie can repeat this procedure, using a new random key bit for
each message bit.

\begin{figure}
  \begin{center}
  \includegraphics[scale=.85]{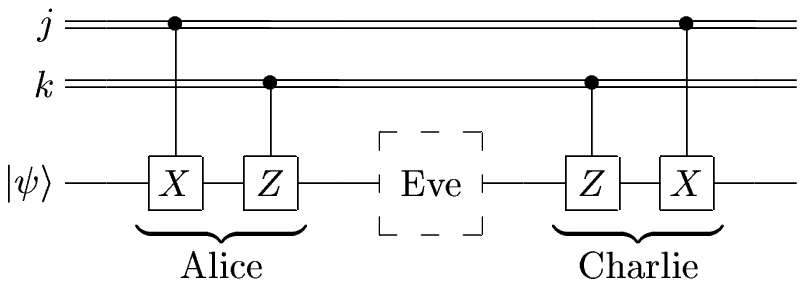}
  \end{center}
  \fcaption{Quantum circuit for the private quantum channel.  The
  double lines represent randomly chosen classical bits $j$ and $k$
  that are shared by Alice and Charlie.  Whether the gates are
  performed or not depends on the values of $j$ and $k$ as indicated.}
  \label{fig:pqc}
\end{figure}

The {\em private quantum channel} is a quantum analogue of this
protocol in which the key remains classical, but the channel is used
to send quantum states \cite{AMTW00,BR00}.  (Note that this differs
from the {\em quantum Vernam cipher}, in which the key is also a
quantum state \cite{Leu01}.)  In the private quantum channel, Alice and
Charlie need to share {\em two} classical bits $j$ and $k$ for Alice
to send her qubit.  The circuit shown in Fig.~\ref{fig:pqc} summarizes
their protocol.  Alice applies the unitary operator $Z^k X^j$ to her
state $|\psi\>$ and sends the result to Charlie.  In between, Eve may
intercept the state, but since she doesn't know $j$ or $k$, she sees
the density matrix
\be
  {1 \over 4} \sum_{j,k=0}^1 Z^k X^j |\psi\>\<\psi| X^j Z^k
  = {I \over 2}
  \label{eq:pqc}
\ee
independent of $|\psi\>$.  From Eve's perspective, Alice has applied
the depolarizing channel, so Eve can learn nothing about the state.
Although she can destroy the state or change it in some way, she
cannot learn anything about it.  Assuming she does nothing, Charlie
will receive the state $Z^k X^j |\psi\>$.  Since he knows the values
of $j$ and $k$, he can apply the inverse operation $X^j Z^k$ to
recover the original state.  If Alice wants to send Charlie $n$
qubits, they can repeat the procedure independently for each qubit,
using a total of $2n$ random classical bits as the key.  From Eve's
perspective, the density matrix of all $n$ qubits is again maximally
mixed, independent of Alice's state, so the procedure is secure.

We will show how the idea of a private quantum channel can be adapted
to allow Bob to help Alice perform a quantum computation.  The main
idea behind these circuits is that Alice can use a private quantum
channel to securely send qubits {\em to herself}.  Since Alice is both
the sender and the receiver, there is no need for her to distribute a
key.  Alice sends her qubits by way of Bob, who plays the part of the
eavesdropper.  However, instead of trying to learn Alice's state
(which of course would be futile), he intentionally performs the gate
Alice would like to be able to do.  After he gives it back to Alice,
she performs an appropriate decoding operation.  Of course, this
procedure is only useful if the decoding operation can be performed
using only the restricted set of gates available to Alice, which is
not necessarily the case for arbitrary gates.

We begin in Sec.~\ref{sec:measurement} by showing how Bob can help
Alice perform a measurement in the computational basis, and then show
how he can help her complete her gate set to do universal quantum
computation in Sec.~\ref{sec:gates}.  We give a unified description of
$k$-round protocols for secure assisted gates in
Sec.~\ref{sec:hierarchy}, and we discuss why they have no meaningful
classical analogue in Sec.~\ref{sec:classical}.  In
Sec.~\ref{sec:honest} we discuss the question of whether Alice can
determine whether Bob is being honest.  Finally, in
Sec.~\ref{sec:discussion} we conclude with a discussion and some open
questions.

\section{Secure assisted measurement}
\label{sec:measurement}

First, we describe how Bob can make a measurement for Alice.  We begin
by discussing a classical version of the protocol.  Suppose Alice has
a classical bit $b$, but although she can manipulate it, she cannot
read its value.  However, she can securely send her bit to Bob and ask
him to read it for her.  Alice chooses a key bit $k$ at random and
computes $b \oplus k$.  She then gives the result to Bob, who reads
the result and tells it to her.  To determine the value of her
original bit $b$, she simply flips the result if $k=1$ and does
nothing otherwise.

\begin{figure}
  \begin{center}
  \includegraphics[scale=.85]{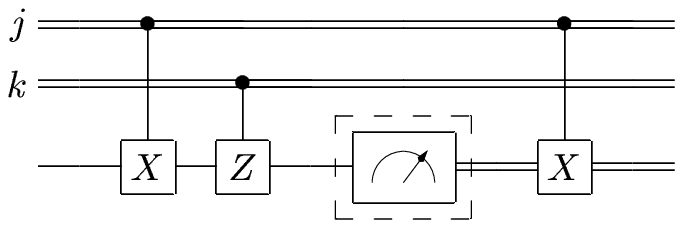}
  \end{center}
  \fcaption{Secure assisted computational basis measurement.  The
  classical bits $j$ and $k$ are randomly chosen, and their values
  determine whether Alice applies certain gates.  The meter inside a
  dashed box represents a computational basis measurement, the action
  performed by Bob if he is honest.  Even if he does some other
  operation, he can learn nothing about Alice's data.}
  \label{fig:measure}
\end{figure}

This classical procedure doesn't seem very useful since reading the
value of a classical bit is usually an easy thing to do.  But quantum
measurement is a difficult task, so we can imagine a scenario in which
Alice can coherently manipulate her qubits, but she cannot measure
them.  In such a situation, there is a quantum version of the above
measurement protocol that allows Bob to make a measurement of Alice's
state in the computational basis.  Alice chooses two random bits $j$
and $k$ and applies the unitary operator $Z^k X^j$ to her state.  She
then gives the qubit to Bob.  He can acquire no information from this
state since from his point of view, by (\ref{eq:pqc}) the density
matrix is maximally mixed, independent of Alice's actual state.
However, if he measures the qubit in the computational basis and
reports the result to Alice, she can determine the result of the
corresponding measurement on her original state.  The $Z$ operator
does not change the measurement result, and the $X$ operator flips it,
so Alice should flip the result if $j=1$ and not flip it if $j=0$.  A
quantum circuit for this protocol is shown in Fig.~\ref{fig:measure}.

\section{Secure assisted gates}
\label{sec:gates}

We now describe how Bob can help Alice perform universal quantum
computation.  We do this by showing that she can perform a universal
set of gates.  In particular, we present circuits by which she can
securely perform the Hadamard gate,
\be
  H := {1 \over \sqrt{2}}\left[\matrix{1 & 1 \cr 1 & -1}\right]\,;
\ee
the $\pi/8$ gate,
\be
  T := \left[\matrix{1 & 0 \cr 0 & \sqrt i}\right]\,;
\ee
and the controlled-{\sc not} gate (a two-qubit interaction),
\be
  C := \left[\matrix{1 & 0 & 0 & 0 \cr
                     0 & 1 & 0 & 0 \cr
                     0 & 0 & 0 & 1 \cr
                     0 & 0 & 1 & 0}\right]
\,.
\ee
This gate set is universal for quantum computation in the sense that
any unitary transformation can be approximated arbitrarily closely by
some sequence of these gates \cite{BMPRV00}.  Thus, the circuits for
secure implementation of these gates can be used as subroutines to
perform an arbitrary quantum computation.

\begin{figure}
  \begin{center}
  \includegraphics[scale=.85]{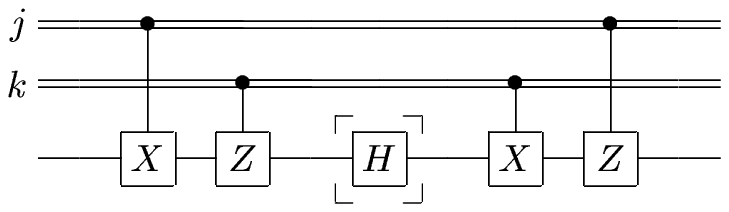}
  \end{center}
  \fcaption{Secure assisted Hadamard gate.  The gate in the dashed box
  is what would be performed by an honest Bob, but if he performs some
  other operation, he will learn nothing about Alice's data.}
  \label{fig:hadamard}
\end{figure}

The simplest of these constructions is the protocol for a secure
Hadamard gate.  Alice chooses two random classical bits $j$ and $k$.
To randomize her state so that Bob can learn nothing from it, Alice
applies $Z^k X^j$ to her qubit.  She then passes it to Bob.  By
(\ref{eq:pqc}), Bob's density matrix is maximally mixed, independent
of Alice's actual state.  If Bob is honest, he performs a Hadamard
gate and hands the qubit back to Alice.  Now she must correct her
qubit so it is as if only the Hadamard were applied.  Because
$XHZ=ZHX=H$, $Z$ can be undone by $X$ and $X$ can be undone by $Z$.
Thus Alice can appropriately fix her state using only Pauli gates,
regardless of the values of $j$ and $k$.  The resulting circuit, shown
in Fig.~\ref{fig:hadamard}, is equivalent to a Hadamard gate if Bob is
honest.  If he is dishonest, he can destroy Alice's qubit or give her
the wrong result, but he can learn nothing from the state she gave
him.

\begin{figure}
  \begin{center}
  \includegraphics[scale=.85]{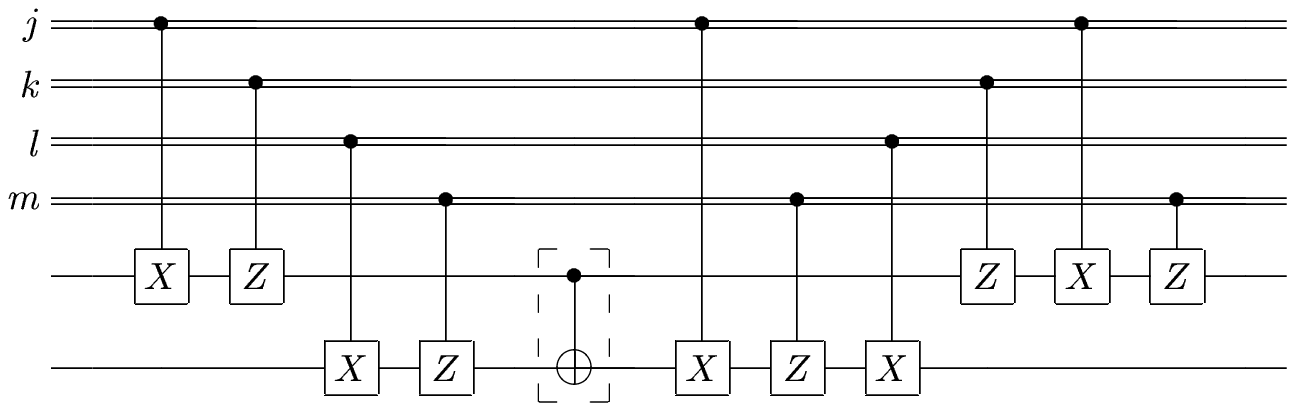}
  \end{center}
  \fcaption{Secure assisted controlled-{\sc not} gate.}
  \label{fig:cnot}
\end{figure}

A similar procedure can be used to perform a controlled-{\sc not}
gate.  Since this is a two qubit gate, Alice must choose {\em four}
random classical bits $j,k,l,m$.  She randomizes her state by applying
$Z^k X^j$ to the first qubit and $Z^m X^l$ to the second.  Then she
gives the qubits to Bob, who is supposed to perform a controlled-{\sc
not} gate and return them to Alice.  Note that by (\ref{eq:pqc})
applied to each of the two qubits, Bob's density matrix is maximally
mixed, independent of Alice's state.  Supposing that Bob performs the
controlled-{\sc not} gate as requested, Alice must correct the encoded
qubits so that the overall interaction is a controlled-{\sc not}.  If
$j=1$, then the target bit was inverted based on an inverted control
bit, so she must apply $X^j$ to the target.  She then fixes the target
bit by applying $X^l Z^m$ and the control bit by applying $X^j Z^k$.
However, if $m=1$, she has also performed a controlled-$(-1)$ gate due
to the anticommutation of $X$ and $Z$.  This can be fixed by applying
$Z^m$ to the control bit.  The resulting circuit is shown in
Fig.~\ref{fig:cnot}.

\begin{figure}
  \begin{center}
  \includegraphics[scale=.85]{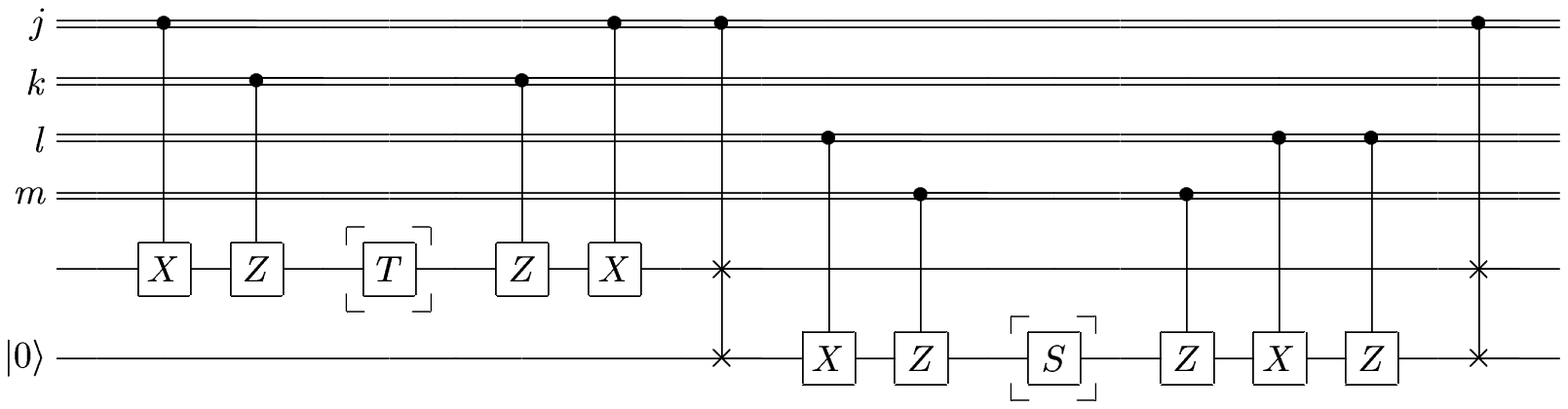}
  \end{center}
  \fcaption{Secure assisted $\pi/8$ gate.  A vertical line with two
  $\times$ symbols indicates a {\sc swap} gate, classically controlled
  on the value of $j$ as indicated.}
\label{fig:pi8}
\end{figure}

Although the $\pi/8$ gate is only a one-qubit gate, this operation is
more complicated to implement: Alice and Bob must use a two-round
protocol, as we prove in the next section.  In total, Alice needs four
random classical bits $j,k,l,m$.  First, she randomizes her quantum
state by applying $Z^k X^j$.  She then gives it to Bob (whose density
matrix, again by (\ref{eq:pqc}), is maximally mixed), and if he is
honest, he applies a $\pi/8$ gate and gives it back to Alice.  She can
undo her randomization by applying $X^j Z^k$.  The $Z$ operation
commutes with $T$, so it does not create any problems.  However, $X T
X = T^\dag$ (up to an overall phase), which differs from $T$ by
$S:=T^2$, a gate that Alice cannot perform.  But she can have Bob do
it for her, again encoding the qubit using (\ref{eq:pqc}), and since
$S^2 = Z$, she will be able to undo the randomization herself.  If she
only asks Bob to help her perform $S$ when $j=1$, this tells him the
value of $j$.  But Alice can avoid revealing $j$ by {\em always}
asking Bob to participate in the second round, but operating on a
dummy qubit when $j=0$.  The complete circuit is shown in
Fig.~\ref{fig:pi8}.  Assuming Bob is honest, this circuit is
equivalent to the $\pi/8$ gate up to an irrelevant overall phase.

\section{The Gottesman-Chuang hierarchy}
\label{sec:hierarchy}

In this section, we explain the existence of the gate constructions
discussed in the previous section using a hierarchy of gates presented
by Gottesman and Chuang \cite{GC99}.  We show that the gates that can
be realized using a $k$-round protocol are exactly those in the
$(k+1)$th level of this hierarchy.

A single-round protocol for a secure assisted gate $U$ on $n$ qubits
requires that
\be
   D_j U E_j = U  \quad\forall~j
\,,
\label{eq:oneround}
\ee
where $E_j,D_j$ are Alice's encoding and decoding operators.  Let
$C_1$ denote the Pauli group on $n$ qubits; that is, all operators
that are tensor products of $n$ Pauli operators.  Furthermore, let
$C_1^S$ denote those operators that can be written as a product of
Pauli group elements and permutations of the qubits (i.e., products of
{\sc swap} gates).  Gates in $C_1^S$ are exactly the operations that
Alice can perform without assistance, so $E_j,D_j \in C_1^S
~\forall~j$.  For simplicity, we will assume that Alice uses the
encoding where $E_j$ runs over all elements of $C_1$, each with
probability $p_j=2^{-n}$.  (More generally, Alice could choose to
include {\sc swap} gates in her encoding, but this would not affect
our conclusions.) The decoding operations are specified by
\be
  D_j = U E_j^\dag U^\dag
\,.
\ee
Thus the requirement $D_j \in C_1^S$ allows us to classify the
possible $U$'s that can be realized in this way: $U$ must come from
the set
\be
  C_2^S := \{U:~U C_1 U^\dag \subseteq C_1^S\}
\,.
\ee
It is not hard to show that in fact\footnote{I thank Wim van Dam for a
discussion of this point.}
\be
  C_2^S = C_2 := \{U:~U C_1 U^\dag \subseteq C_1\}
\,.
\ee
The set $C_2$ forms a well known group, the {\em Clifford group}
\cite{Got99}.  Any gates in this group, such as the Hadamard and {\sc
cnot} gates, can be realized with a one-round protocol.

If Alice and Bob are willing to use a two-round protocol, they can do
more gates.  By the preceding discussion, Bob must do a gate from
$C_2$ in the second round.  However, in the first round, he can do any
gate from
\be
  C_3 := \{U:~U C_1 U^\dag \subseteq C_2\}
\,,
\ee
e.g., the $\pi/8$ gate (or the Toffoli or Fredkin gate).  Different
encoding/decoding pairs may require Bob to do different things in the
second round, but if Alice uses the {\sc swap} trick introduced in the
secure $\pi/8$ gate construction, this need not provide him with any
information.  By induction, we see that in $k$ rounds, Alice and Bob
can securely perform any gate in the set
\be
  C_k := \{U:~U C_1 U^\dag \subseteq C_{k-1}\}
\,.
\ee
Note that gates in $C_2$ are not sufficient for universal computation.
However, universal quantum computation can be done using $C_3$ gates.

At first glance, it may seem that the form of (\ref{eq:oneround}) is
too restrictive.  Why should Bob have to perform exactly the gate
Alice wants?  In other words, shouldn't we consider protocols of the
form
\be
  D_j V E_j = U \quad \forall~j
\,,
\label{eq:bobdifferent}
\ee
where $V \ne U$?  The answer is no, because if such a scheme exists,
we can easily turn it into a scheme where $V=U$: we have $V = D_0^\dag
U E_0^\dag$, which can be substituted into (\ref{eq:bobdifferent}) to
give
\be
  U = D_j D_0^\dag U E_0^\dag E_j
\,.
\ee
By using the modified encoding operations $E_j'=E_0^\dag E_j$ and
decoding operations $D_j' = D_j D_0^\dag$, Alice can ask Bob to
perform $U$ instead of $V$.  Thus there is no loss of generality in
the choice (\ref{eq:oneround}).

\section{Secure assisted classical gates}
\label{sec:classical}

It is interesting to contrast these protocols for secure assisted
quantum gates with the corresponding problem for reversible classical
computers.  If Alice cannot do universal classical computation, is it
possible for Bob to assist her in a secure way?  The answer is no:
secure assisted universality can only be achieved in a meaningful way
in the quantum setting.  This is essentially because the ability to do
classical operations is assumed in the quantum case, so classically
controlled quantum gates are no more difficult to implement than
unconditional quantum gates; but a classically controlled classical
gate can be a more powerful resource than the original classical gate.
For example, the controlled-{\sc not} gate is not universal for
classical computation, but a controlled-controlled-{\sc not} (more
commonly known as a Toffoli gate) {\em is} universal.

To see why secure assisted classical universality is not possible, we
introduce a classical analogue of the Gottesman-Chuang hierarchy.  For
$n$-bit gates, let $\tilde C_1$ denote gates that are tensor products
of $n$ identity and {\sc not} gates.  Also, let
\be
  \tilde C_k := \{P:~P \tilde C_1 P^{-1} \subseteq \tilde C_{k-1}\}
\ee
denote the set of classical gates (permutations) that map $\tilde C_1$
gates into $\tilde C_{k-1}$ gates under conjugation.  Just as in the
quantum case, gates in $\tilde C_2$ are not sufficient for universal
computation, but there are gates from $\tilde C_3$ (e.g.,\ the Toffoli
and Fredkin gates) that are universal.  Thus the natural restriction
on Alice's gate set is to allow her to do only $\tilde C_2$ gates.  In
particular, since she must use gates that are controlled based on the
values of her random key bits, she will be interested in using
controlled-$\tilde C_1$ gates, all of which are in $\tilde C_2$.
However, with a single-round protocol, she can only use
controlled-$\tilde C_1$ gates to build $\tilde C_2$ gates on her
computational bits, which gives her no additional computational power.
Since she cannot perform the controlled-{\sc swap} (Fredkin) gate, she
cannot securely perform a multi-round protocol.  Thus there is no way
for her to achieve secure assisted universality.

However, there are particular examples in which Alice and Bob can
perform a computation securely even in the classical case.  Some work
along these lines was mentioned in Section~\ref{sec:intro}, but a
simple example that shows the advantage of having a particular problem
in mind is the following.  Suppose Alice would like to find a
satisfying assignment for a Boolean formula containing $n$ variables.
Alice generates $n$ random bits, one for each of the variables in the
formula.  If the bit corresponding to a particular variable is zero,
she does nothing; if that bit is one, she inverts the variable
wherever it appears in the formula.  She then tells Bob the formula,
and if he is honest, he gives her a satisfying assignment.  To find a
satisfying assignment for her original problem, she flips the bits
corresponding to the variables that were inverted in the formula she
gave to Bob.  Although Bob can learn a lot about the structure of the
problem, he cannot learn the particular satisfying assignment of her
original problem.\footnote{I thank Sam Gutmann for suggesting this
example.}

\section{Keeping Bob honest}
\label{sec:honest}

The primary weakness of these protocols is that although he can learn
nothing about the states Alice gives him, Bob can easily prevent her
from performing her computation.  He could simply not return her
qubits, or worse yet, he could ruin the computation by performing the
wrong gate.  This weakness is inherited from the private quantum
channel, and there is no way to avoid it altogether.  However,
although she cannot force Bob to help her, there may be simple ways
for Alice to detect whether Bob is cheating.

If Alice is asking Bob to help her solve a problem in the
computational complexity class NP, there is a particularly simple way
for her to check his honesty.  Presumably, Alice is asking for Bob's
help because the problem can be solved much faster on a quantum
computer than on a classical computer---for example, she might ask him
to help her perform Shor's algorithm to factor a number.  But if Alice
has access to a classical computer and the problem is in NP, she can
easily check the solution to see if it is right.  In the example of
factoring, Alice simply multiplies the resulting factors to see if
they give her input.

However, what if Alice cannot readily check the solution?  Can she
still efficiently detect whether Bob is cheating?  Intuitively, it
seems that Alice should be able to gain some confidence that Bob is
behaving honestly by performing tests of his actions using a randomly
chosen subset of her inputs.  If these tests fully characterize Bob's
operations, they can be used to bound the probability that Bob is
cheating.  Indeed, such a procedure provides an efficient way to check
whether Bob is cheating in the restricted scenario in which he must
act as a memoryless black box, as can be shown using ideas along the
lines of \cite{DMMS00}.  However, in general, Bob could introduce
errors adversarially rather than randomly, which presents a more
difficult verification problem.  We leave the general adversarial
scenario as an open problem.  A solution to this problem might also be
relevant to fault-tolerant quantum computing, where the assumption
that Bob acts as a memoryless black box corresponds to an assumption
of independent errors, and the general adversarial scenario
accommodates errors of the most general type (which are hopefully
nevertheless small).

\section{Discussion}
\label{sec:discussion}

We have shown that it is possible for a party who cannot do universal
quantum computation (Alice) to have her computational power augmented
by another party (Bob) without compromising the security of the
computation.  Furthermore, we have briefly discussed ways of detecting
whether Bob is truly being helpful---a problem that deserves further
study.  The protocols we have described, and more general protocols
for verifying the validity of Bob's actions, might prove useful for
assuring the security of certain quantum information processing tasks.

We have focused on preventing Bob from obtaining information about the
states Alice gives him, and we have not considered the information he
might obtain from the particular gates Alice asks him to perform.  We
can imagine that she might want to prevent him from learning something
this way.  For example, in the context of programmable gates, Vidal
and Cirac considered a different scenario in which Bob can learn
Alice's input, but she does not want him to know the function she is
trying to compute \cite{VC00}.  In the context of the present paper,
it is not particularly difficult to prevent Bob from learning the
function.  The protocol can simply consist of Bob performing a fixed
sequence of gates, cycling through Hadamard, {\sc cnot}, and $\pi/8$.
If a particular gate is not needed, she can supply Bob with junk
qubits.  With this protocol, the number of gates is increased by at
most a factor of three.  Since Alice does not send any classical
information to Bob to describe her circuit, and since we have already
established that he can learn nothing from her quantum states, it
follows that he cannot learn anything about which gates are being
used.  The only thing Bob can learn is the length of the protocol,
i.e., the total number of gates Alice has him perform.  Even this
meager amount of information can be reduced (although not eliminated),
since Alice is free to add additional unnecessary gate requests, so
that Bob can only learn an upper bound on the number of gates in
Alice's circuit.

Note that there is an analogy between programmable gates and secure
assisted gates: whereas programmable gates generalize identity
teleportation to gate teleportation, secure assisted gates generalize
an identity private quantum channel to a private quantum channel that
performs a gate.  Furthermore, the Gottesman-Chuang hierarchy plays a
similar role in the construction of gate teleportation circuits
\cite{ZLC01} and in showing that two-qubit measurements are universal
for quantum computation \cite{Leu01a}.

There are many possible variants of this problem depending on the
resources Alice and Bob are allowed to use.  For example, if Alice is
only allowed to perform single-qubit measurements, Bob can supply her
with a cluster state \cite{RB01}.  Perhaps other resource limitations
would lead to interesting forms of secure assisted quantum
computation.

Finally, it might be useful to consider restricting the total amount
of information transfer.  We have assumed that Alice and Bob have an
inexpensive quantum channel, so they can send quantum states back and
forth as many times as they wish.  But this may not be a realistic
assumption.  If Alice and Bob are connected only by a very slow or
expensive channel---or perhaps only by a classical channel, with a
small reserve of prior shared entanglement---can they still accomplish
interesting computational tasks?  In other words, can they perform
{\em secure remote quantum computation}?  We should not expect Bob to
enable Alice to do secure universal computation on remote data, but
she might nevertheless be able to perform certain tasks securely.


\nonumsection{Acknowledgements}
\noindent
I thank Isaac Chuang, Wim van Dam, Edward Farhi, Jeffrey Goldstone,
Sam Gutmann, Aram Harrow, Debbie Leung, Hoi-Kwong Lo, Ben Recht, Adam
Smith, and Ronald de Wolf for helpful discussions.  This work was
supported by the Fannie and John Hertz Foundation, by the
Cambridge--MIT Institute, by the National Security Agency and Advanced
Research and Development Activity under Army Research Office contract
number DAAD19-01-1-0656, and by the Department of Energy under
cooperative research agreement DE-FC02-94ER40818.


\nonumsection{References}

\end{document}